\documentclass[%
aip,jcp,amsmath,amssymb, linenumbers, reprint
]{revtex4-2}
\usepackage{graphicx}
\usepackage{dcolumn}
\usepackage{bm}
\usepackage{amssymb}
\usepackage{amsmath}
\usepackage{supertabular}
\usepackage[table,x11names]{xcolor}
\usepackage{longtable}
\usepackage{colortbl}
\usepackage{multirow}\usepackage{braket}
\usepackage{esvect}
\usepackage{gensymb}
\usepackage{soul}
\let\oldAA\AA
\renewcommand{\AA}{\text{\normalfont\oldAA}}

\usepackage{qtree}
\usepackage{tipa}
\usepackage{appendix}
\usepackage{ulem}
\usepackage{xcolor}
\usepackage{tabularx,booktabs}
\usepackage{simplewick}
\usepackage{simpler-wick}
\usepackage{multirow, makecell}
\usepackage{scalerel}

\usepackage[colorinlistoftodos]{todonotes}
\presetkeys{todonotes}{inline,backgroundcolor=white}{}

\begin{document}
\nolinenumbers
\title{Dual Exponential Coupled Cluster Theory: Unitary Adaptation, Implementation in the Variational Quantum Eigensolver Framework and Pilot Applications}
\author{Dipanjali Halder$^{1}$, V. S. Prasannaa$^{2}$, Rahul Maitra$^{1,\dagger}$}
\affiliation{$^1$ Department of Chemistry, \\ Indian Institute of Technology Bombay, Powai, Mumbai 400076, India \\
$^2$ Centre for Quantum Engineering, Research and Education, TCG CREST, Salt Lake, Kolkata 700091, India\\
$^\dagger$ rmaitra@chem.iitb.ac.in}

\date{\today}

\begin {abstract}
In this paper, we have developed a unitary variant of a double exponential coupled cluster 
theory, which is capable of mimicking the effects of connected
excitations of arbitrarily high rank, using only  rank-one and rank-two parametrization of the wavefunction ans\"{a}tz. 
While its implementation in a classical computer necessitates the construction of an 
effective Hamiltonian which involves infinite number of terms with arbitrarily high 
many-body rank, the same can easily be implemented in the hybrid quantum-classical 
variational quantum eigensolver framework with a reasonably shallow quantum circuit.
The method relies upon the nontrivial action of a unitary, containing a set of rank-two
scattering operators, on entangled states generated via cluster operators. We have 
further introduced a number of variants of the ans\"{a}tz with different degrees of 
expressibility by judiciously approximating the scattering operators. With a number of applications on strongly correlated molecules, we have shown that all our schemes can 
perform uniformly well throughout the molecular potential energy surface without 
significant additional implementation cost and quantum complexity over the unitary 
coupled cluster approach with single and double excitations.
\end{abstract}

\maketitle

\section{Introduction}\label{sec1}
The emerging field of quantum computation has gained substantial attention due to its promise to solve certain computational problems that are difficult to handle with classical computers~\cite{deutsch, shor, Ortiz, hhl}. Simulation of many body quantum systems is considered to be one such strenuous task owing to the exponential overhead of computational resources, thus limiting its application to small chemical systems. Unlike exact classical methods like full configuration interaction (FCI) that scales exponentially with system size, quantum algorithms such as quantum phase estimation (QPE) can tackle such problems with polynomial overhead. An overview of the developments in this field can be found in relevant review articles ~\cite{ag1, ag2}.

Earlier quantum computational algorithms to simulate molecular energies were based on the QPE  approach as envisioned by Abrams and Lloyd~\cite{Abrams_and_Lloyd_1997,Abrams_and_Lloyd_1999}. The objective of the phase estimation approach is to extract eigenvalues of a unitary operator by projecting the input state onto an eigenstate, provided the input state has substantial overlap with the exact eigenstate of the unitary operator. In 2005, Aspuru Guzik \textit{et al.}~\cite{Aspuru_Guzik_2005} first employed the phase estimation algorithm to simulate ground state energies of a few molecular systems. Eventually, Hefeng Wang and co-workers estimated excited state energies employing multiconfigurational self-consistent field (MCSCF) wavefunction in the framework of QPE~\cite{Hefeng_Wang_2008}. Apart from these papers, there have been numerous works in the QPE literature which deal with techniques for preparation of initial states suitable for the phase estimation approach, for example, see Ref.~\cite{Pittner, (doi/10.1021/acscentsci.8b00788), iqpeovervqe}. Although it offers an exponential speedup over the exact classical algorithms, QPE fails to meet the expectations of the near term noisy intermediate scale quantum (NISQ) devices owing to the requirement of long coherence times. 

To circumvent this problem, a hybrid classical-quantum approach, 
namely variational quantum eigensolver (VQE) was proposed by 
Peruzzo and co-workers~\cite{Peruzzo_2014}. The idea stems from
the Rayleigh-Ritz variational principle~\cite{Griffiths}, which dates back to early 
1900s. VQE utilizes a combination of both
classical and quantum architectures to find the best variational
approximation to the ground state of a given Hamiltonian
corresponding to a chosen trial wavefunction ans\"{a}tz. Till date, VQE has been
experimentally implemented in various quantum hardware
architectures, viz. photonic quantum 
processors~\cite{Peruzzo_2014}, 
superconducting quantum processors~\cite{Malley, Colles}, 
trapped ion architectures~\cite{Shen, Hempel_trapped_ion, Kokail} etc. It is worth 
mentioning at this point that the selection of a suitable 
wavefunction ans\"{a}tz plays a pivotal role in the performance
of VQE. In recent times, there have been a number of developments
to formulate an expressive and cost efficient ans\"{a}tz for VQE. An extensively used ans\"{a}tz is the chemically
motivated unitary coupled cluster ans\"{a}tz, abbreviated as UCC~\cite{ucc, Yung, Peruzzo_2014,  evangelista, Romero_2018, Anand}. A generalized extension of the
unitary coupled cluster ans\"{a}tz truncated at singles and 
doubles excitations (UCCGSD), inspired from the earlier works 
of Nakatsuji~\cite{nakatsuji} and Nooijen~\cite{nooijen}, has been studied by Van Voorhis and Head-Gordon~\cite{Voorhis}, and further implemented in the VQE framework by Lee \textit{et 
al.}~\cite{head_gordon}. UCCGSD is considered to be the most expressive ans\"{a}tz
to date, but the expressibility comes at the expense of much higher scaling in terms of
the number of parameters and the circuit depth (in comparison to
UCCSD). In the same paper, Lee and co-workers have further 
proposed an ans\"{a}tz, termed as UpCCGSD, which is analogous 
to the UCCGSD ans\"{a}tz, except for an added constraint of
inclusion of \textit{only generalized paired excitations} in 
the definition of two-body operators. UpCCGSD has been found to 
scale linearly against the system size, thus leading to a synergy
between efficiency (in terms of cost of implementation), 
expressibility and the accuracy. The authors further showed that 
one may reach chemical accuracy across the potential
energy surface of strongly correlated molecules with UpCCGSD 
ans\"{a}tz by repeating the UpCCGSD circuit $k$ number of times.  
This leads to $k$-fold increase of the circuit depth which 
anyway scales linearly against the system size. The same train of thought has led to the development of multireference unitary coupled cluster with partially generalized singles and doubles (MR-UCCpGSD) ans\"{a}tz~\cite{MR-UCCpGSD} by Sugisaki and co-workers. Based on a
slightly different idea, Grimsley \textit{et al.}~\cite{ADAPT-VQE} introduced a systematically expandable  ans\"{a}tz, the adaptive derivative-assembled pseudo-Trotter variational quantum eigensolver, abbreviated as ADAPT-VQE. The underlying idea behind such an ans\"{a}tz is to extract most of the correlation energy with the lowest possible number of variational parameters (or fermionic excitation operators). 
A complementary approach for enhancing  the efficiency (from the quantum complexity reduction perspective) of the VQE algorithm involves the implementation of a downfolded Hamiltonian~\cite{ducc, ducc1} or a rotated Hamiltonian~\cite{oo-ucc1, oo-ucc2}. Few more notable works in this direction includes Unitary Cluster Jastrow (uCJ) ans\"{a}tz~\cite{uCJ}, low rank representations~\cite{LR}, which are known to have a quadratic scaling of the parameters against the system size.  
Apart from the aforementioned developments in the context of traditional VQE, Stair \textit{et. al.}~\cite{PQE} introduced a variant of VQE, namely Projective Eigensolver algorithm (abbreviated as PQE). Based on the projective techniques used for solving traditional coupled cluster equations, PQE is known to carry out optimization through the residue calculations (on a quantum hardware), unlike the energy gradient calculations employed in traditional VQE.  

While the success of VQE depends on the choice
of parameters in the UCC ans\"{a}tz, in recent times, there is an increasing interest in the 
many-body electronic structure community to explore the 
possibility of writing an $N-$electron correlated 
wavefunction with lower rank parametrization. One of the present authors had worked in this direction, and introduced a double exponential
waveoperator based coupled cluster theory, which implicitly
incorporates the effects of connected rank-three and rank-four 
excitations with only rank-one and -two parametrization, 
in a computationally inexpensive manner~\cite{RM, Anish}. In the said theory, 
termed as iterative n-body excitation inclusive CCSD (iCCSDn),
the authors had introduced a set of rank-two scattering 
operators on top of the conventional single and double
excitation operators. iCCSDn relies heavily on 
the non-commutativity of these two sets of operators,
and the connected high rank excitations are simulated through
the tensor contractions between them. 

With the advent of VQE algorithm for the calculations
of atomic and molecular ground state energetics with a variety
of unitary CC ans\"{a}tze, it is compelling to verify the efficacy
and accuracy of the iCCSDn methodology on a quantum architecture.
As the first step 
towards achieving this, we develop a unitarized version of 
iCCSDn (UiCCSDn) theory and a number of variants thereof, 
and implement those on a quantum 
simulator for the molecular ground state energy evaluation.
The various UiCCSDn variants seem to be extremely precise 
over the entire range of potential energy surface without a significant increase in the gate count as compared to the conventional UCCSD. This
point will be elaborated towards the end of the theory section
and also in the results and discussion 
section in the later part of the manuscript. 

The paper is organized as follows: we discuss in section
~\ref{theory} the general VQE algorithm, with emphasis on the
chemically motivated unitary (generalized) coupled cluster
ans\"{a}tz. This would be followed by a brief discussion on
the conventional double exponential ans\"{a}tz, iCCSDn, in 
section ~\ref{conv iCCSDn}. In section ~\ref{UiCCSDn}, we
introduce the unitarized version of the aforementioned ans\"{a}tz
and its various approximations along with their efficiency of
implementation in the context of VQE. We will compare
and contrast various aspects of (U)iCCSDn in its classical and
quantum implementation. Finally, we present our results in 
section ~\ref{results}, where we would study the potential 
energy surface of a few strongly correlated molecular systems
and demonstrate that our methods can handle electronic states
with varied complexity with uniform accuracy. Finally, we
conclude in section \ref{conclusion} along with a road-map to 
future directions. 

\section{Theory}\label{theory}

\subsection{The variational quantum eigensolver algorithm with unitary (generalized) coupled cluster ans\"{a}tz}

The hybrid classical-quantum VQE algorithm uses a combination of both classical and quantum computation to find an upper bound to the ground state energy associated with a given 
Hamiltonian, a problem central to the field of quantum chemistry. It is based on the 
variational principle, and hence the ground state energy functional can be written as,
\begin{eqnarray}
E(\theta) &=& \frac{\langle \Psi(\theta)|H|\Psi(\theta) \rangle}{\langle \Psi(\theta)|\Psi(\theta) \rangle} \nonumber \\
&=& \langle \Psi_{HF}|U(\theta)^\dag HU(\theta)|\Psi_{HF} \rangle, \label{eqE} 
\end{eqnarray}
where $|\Psi(\theta) \rangle$ is a parameterized trial wavefunction, $|\Psi_{HF} \rangle$ is a suitable many-body reference state, which is usually the Hartree-Fock (HF) state, and H is the electronic Hamiltonian.
The algorithm is divided into two subparts:
the first subpart uses a quantum computer for preparation of a parameterized wavefunction or 
the ans\"{a}tz and measurement of the expectation value of the Hamiltonian. And the second 
subpart comprises of a classical optimization algorithm that minimizes the measured energy 
and updates the variational parameters for ans\"{a}tz preparation. Thus, we can write,
\begin{eqnarray}
E_{0} = \mathrm{min}_{\theta} \langle \Psi_{HF} | U(\theta)^\dag H U(\theta) | \Psi_{HF} \rangle, \label{eqE1}
\end{eqnarray}
where $U(\theta)$ is a unitary parametrized operator. 
The many-body molecular Hamiltonian, H, can be represented in second quantized form as shown in Eq.~(\ref{hamiltonian}),
\begin{eqnarray}
H&=&\sum\limits_{p,q}h_{pq}a_{p}^{\dagger}a_{q}+\frac{1}{2}\sum\limits_{p,q,r,s}h_{pqrs}a_{p}^{\dagger}a_{q}^{\dagger}a_{r}a_{s}.
\label{hamiltonian}
\end{eqnarray}

One of the crucial components of VQE is the choice of the parametrized ans\"{a}tz. 
Till date, various types of ans\"{a}tze have been explored in the VQE literature. A few important things to keep in mind while selecting an ans\"{a}tz are: 
it should be able to span the N-electron Hilbert space,  and the circuit depth
corresponding to the ans\"{a}tz should not be beyond the reach of near term NISQ devices. 
One widely used ans\"{a}tz is the chemically motivated UCC 
ans\"{a}tz. It uses an unitary exponential parametrized wave operator, $e^X$ to span the
$N$-electron Hilbert space. There have been a number of developments along this line to 
judiciously choose $X$. In the traditional UCC method, the operator is 
chosen to be the sum of an excitation operator, X (defined with respect to a 
Fermi vacuum, usually chosen as the HF determinant) and its de-excitation counterpart, $X^{\dag}$. 
The operator $X$ (and $X^\dagger$) is usually truncated with rank-one and rank-two 
operators, giving rise to the well known UCCSD ans\"{a}tz:
On the other hand, the operator $X$ may be chosen to be agnostic to the hole and 
particle orbitals, and one may define $X$ to be an anti-hermitian sum of 
generalized rank-one and rank-two operators. This is known as the UCCGSD ans\"{a}tz in the 
literature.

In the most general form,
\begin{eqnarray}\label{uccop}
\ket{\Psi(\theta)}_{UCC(G)SD}&=&e^{{{X}({\theta})}-{X}^{\dagger}({\theta})}\ket{\Psi_{HF}}; \label{excop} 
\end{eqnarray}
where $X$ is the cluster operator,
\begin{eqnarray}
X &=& X_{1}+X_{2},
\end{eqnarray}
where
\begin{eqnarray}
{X_{1}}({\theta})&=&\sum \limits_{p;q}\theta_{p}^{q}a_{q}^{\dagger}a_{p};\label{t1}\\
{X_{2}}({\theta})&=&\sum \limits_{p,r;q,s}\theta_{pr}^{qs}a_{s}^{\dagger}a_{q}^{\dagger}a_{r}a_{p}. \label{t2}
\end{eqnarray}
Here, the subscripts $p,q,r,s$ refer to general spinorbital indices for UCCGSD. If one 
restricts the orbitals $p,r$ to occupied orbitals $i,j$ and $q,s$ to the unoccupied 
orbitals $a,b$, the UCCGSD ans\"{a}tz reduces to  the UCCSD  one, with $X_1(\theta) \equiv T_1$, $X_2(\theta) \equiv T_2$, and $T = T_1 + T_2$. An efficient variant of the UCCGSD
asnsatz was proposed by Lee \textit{et. al.} where the authors had included only generalized 
pair double excitations where two electrons are ``excited" from a given spatial orbital to 
another spatial orbital. That implies that any combination $p,r$ from Eq. (7) have the same spatial orbital and so do their corresponding
$q,s$. The resulting Up-CCGSD ans\"{a}tz has linear circuit depth with respect to the number 
of spinorbitals, and is thus considered to be the current state of the art. Note that the 
amplitudes, 
$\theta_{p}^{q}$ and $\theta_{pr}^{qs}$ are the VQE parameters, which are iteratively 
updated through an appropriate optimizer.

In the following sections, we first present the parent iCCSDn theory from a many-body theoretic 
perspective and discuss its various advantages and disadvantages in its implementation on a 
classical computer. This would set the stage for the next section,  where we
would develop the unitary version of the theory and show that the unitarized version
is perfectly suited to be implemented on a quantum computer.

\subsection{The double exponential ans\"{a}tz based conventional iCCSDn}\label{conv iCCSDn}

In the conventional iCCSDn theory, one aims to simulate the high rank correlation 
effects (dominantly triples and quadruples) through a sequential similarity transformation
with the exponentiated rank one and rank two operators.
Towards this, one introduces a set of rank-two \textit{scattering} operators, $S$, whose structure 
contains a true excitation in one of the vertices and a hole-hole/particle-particle 
scattering in another vertex. The scattering vertex contains a single (quasi-) hole or 
(quasi-) particle type one-electron state destruction operator. Depending on whether a hole type 
or a particle type of orbital that appears as the destruction operator in the scattering 
vertex of $S$, one may classify it as $S_h$ and $S_p$.
\begin{eqnarray}
S_h &=& \frac{1}{2} \sum_{amij} s^{am}_{ij} \{a^\dagger m^\dagger ji\}; \nonumber \\
%\hspace{0.5cm}
S_p &=& \frac{1}{2} \sum_{abie} s^{ab}_{ie} \{a^\dagger b^\dagger ei\} \mathrm{\ \ and} \nonumber \\
S &=& S_h + S_p. 
\label{eq4}
\end{eqnarray}
Here, \textit{a, b, c, ...} etc. denote the particle orbitals 
and \textit{i, j, k, ...,} etc. are the hole orbitals with 
respect to HF determinant taken as the Fermi vacuum. In the second quantized
representation of the $S$ operators, the hole state $m$ in $S_h$ and
the particle state $e$ in $S_p$ are the ones that appear as the 
destruction operators. Note that in the conventional CC theory with HF determinant
taken as the Fermi vacuum, all the quasi-particle orbitals (e.g. \textit{a, b, c, ...}) and quasi-hole orbitals (e.g. \textit{i, j, k, ...}) appear 
as the creation operators. In order
to distinguish the hole/particle states that appear as the destruction operator
in $S$, we have used different symbols for them.

The $S$ operators, due to the presence of the destruction operators, annihilate the
HF reference determinant: $S|\Psi_{HF}\rangle=0$. This is to be referred as the 
\textit{killer condition}. Furthermore, the $S$ operators do not commute among 
themselves, neither do they commute with the cluster operators $T$. However, they
have non-trivial action on an excited determinant where a given hole or particle 
state is created by the preceding action of an $T$ operator. In fact, the non-commutativity
between the $S$ and the $T$ operators are the key to simulate the connected 
high-rank excitations. In order to construct a terminating series of the effective
Hamiltonian, in the conventional many-body formalism of iCCSDn, the $S$ operators are
not allowed to contract among themselves. 

Due to the killer condition and the non-commutativity of the $S$ and $T$ operators,
one may write down a double exponential parametrized ans\"{a}tz:
\begin{equation}
%\Omega = \{\exp(S)\}\exp(T_1 + T_2)
\Omega = \{e^S\}e^{T_1 + T_2}, 
\label{ansatzeq}
\end{equation}
where the action of the two sets of exponential operators are fixed, This 
is to be distinguished with an ans\"{a}tz like $e^{\{S\}+T}$, where the terms appearing
in the exponent can appear in any order. Note that the $\{...\}$ denotes 
the normal ordering to avoid the $S-S$ contraction to ensure a naturally terminating 
effective Hamiltonian structure. In order to have a non-trivial coupling among $S$ and $T$ operators,
the destruction operator in $S$ should \textit{necessarily} get
contracted with one of the creation operators present in $T$.

The solutions for the optimized $s$ and $t$ amplitudes (corresponding to $S$ and $T$
operators, respectively) are done in a sequentially coupled manner. One defines the 
first similarity transformed effective Hamiltonian, $W$ such that 
\begin{equation}
%\{\exp(S)\} W = H \{\exp(S)\}
\{e^S\} W = H \{e^S\}
\label{eqxx4}
\end{equation}
is satisfied. Since the inverse operation of a normal ordered ans\"{a}tz is not
explicitly defined, one may determine $W$ with a recursive substitution technique. 
The resulting structure of $W$ takes the form:

\begin{eqnarray}
W=\{\contraction{}{H}{}{\exp(S)} H \exp(S)\}  - \{\contraction[2ex]
{}{(\exp(S)-1)}{}{\contraction{}{H}{}{\exp(S)} H \exp(S)} (\exp(S)-1)
\contraction{}{H}{}{\exp(S)} H \exp(S)\} + \nonumber \\
\{\contraction[3ex]{}{(\exp(S)-1)}{}{\contraction[2ex]
{}{(\exp(S)-1)}{}{\contraction{}{H}{}{\exp(S)} H \exp(S)} (\exp(S)-1)
\contraction{}{H}{}{\exp(S)} H \exp(S)} (\exp(S)-1) \contraction[2ex]
{}{(\exp(S)-1)}{}{\contraction{}{H}{}{\exp(S)} H \exp(S)} (\exp(S)-1)
\contraction{}{H}{}{\exp(S)} H \exp(S) \}  - \cdots . 
\label{eqxx6}
\end{eqnarray}

This is an effective operator containing many-body terms, where usually 
only the rank-one and rank-two terms are retained and the series is restricted 
till the second term in the right hand side. With a judiciously truncated $W$, one
may employ the second set of similarity transformation to construct double 
similarity transformed effective Hamiltonian $H^{eff}=e^{-T}We^T$. Note that 
in the double similarity transformed effective Hamiltonian, $H^{eff}$, both the 
$T$ and $S$ operators get coupled. Their amplitudes are determined by 
employing a many-body expansion and demanding that the corresponding amplitudes
of $H^{eff}$ vanish.
\begin{equation}
(h^{eff})^{a}_{i} = (h^{eff})^{ab}_{ij} = (h^{eff})^{am}_{ij} = (h^{eff})^{ab}_{ie} 
= 0. 
\label{eqxx18}
\end{equation}
Note that in the construction of $H^{eff}$, the $s$ and the $t$ amplitudes are 
iteratively optimized in a coupled manner, and as such the high rank correlation 
effects are simulated through the contraction between them: $T_{ijk}^{abc} \leftarrow S_{ij}^{am}T_{mk}^{bc} + S_{ie}^{ab}T_{jk}^{ec}$.

\subsection{Unitary version of double exponential CC ans\"{a}tz}\label{UiCCSDn}

\subsubsection{Motivation towards the development of the unitarized ans\"{a}tz}

While the iCCSDn methodology is shown to be quite accurate, particularly for weakly
to moderately strong correlated systems, there are a few drawbacks that it suffers
from in its implementation on a classical computer.

\begin{figure*}[t]
    \centering
            \includegraphics[height=80mm,width=150mm]{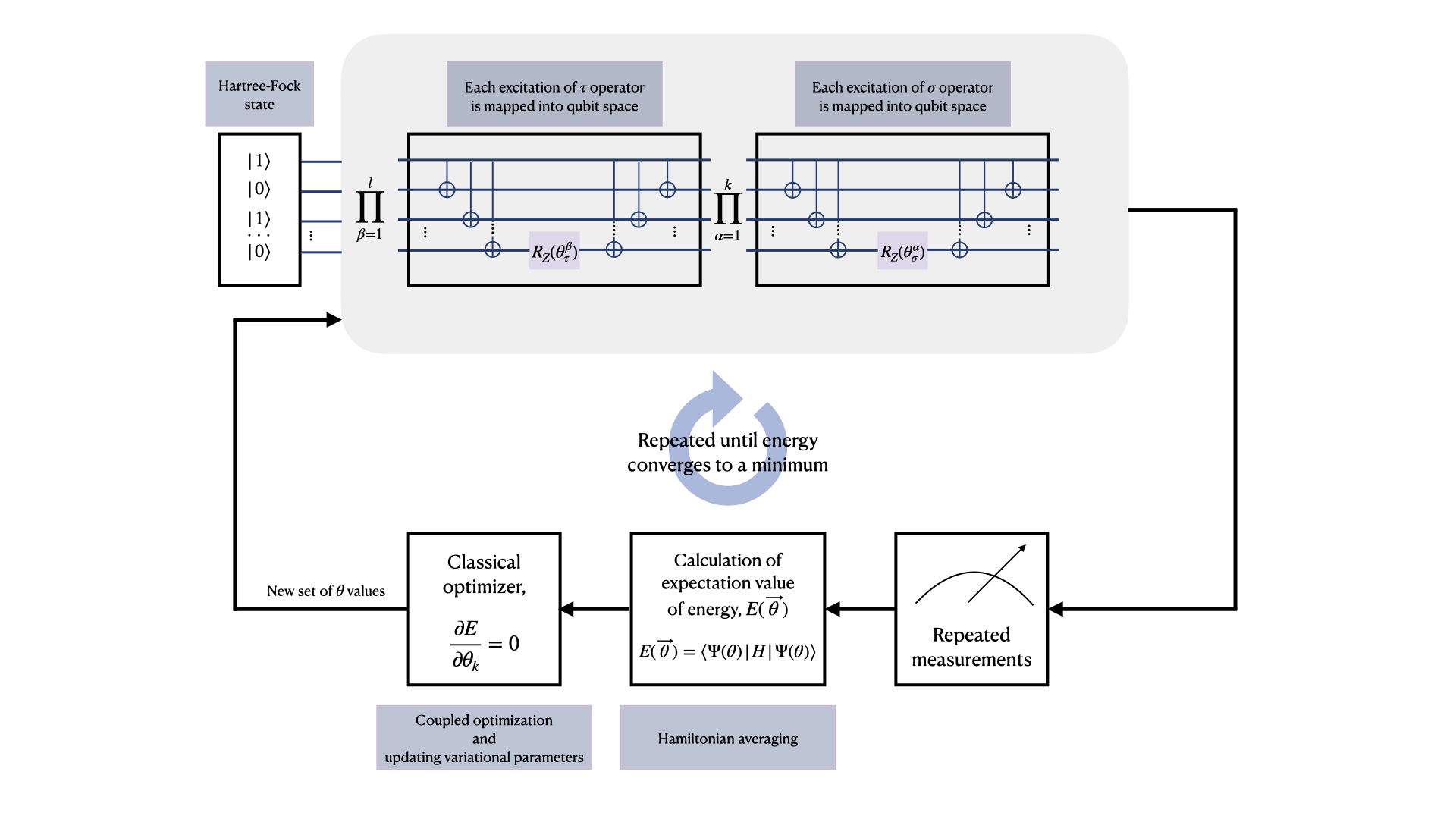} 
\caption{Schematic illustration of the UiCCSDn scheme for an $n$-qubit example system. Note that each of the two circuits shown in the figure is a representative illustration of an excitation (the first block) and scattering (the second block), and does not reflect an actual full circuit, which is much deeper. }
\label{circuit}
\end{figure*}

\begin{enumerate}
    \item The expansion in Eq. (\ref{eqxx6}) (with a normal ordered ans\"{a}tz as shown) is exact only up to
    leading order, but is naturally terminating. However, in the practical implementation, the series is usually further truncated retaining only a subset of the leading terms. That leaves out a large number of terms that otherwise could effectively contribute to the amplitude equations, Eq. (\ref{eqxx18}). Furthermore, the rank of the terms appearing in the effective Hamiltonian, $W$
    is truncated to one and two-body that essentially leaves out the effects of high 
    rank (beyond triples) excitations. Thus the ans\"{a}tz with a normal ordered exponential 
    structure has less \textit{span} of the $N-$electron Hilbert space, compared to the
    one without the normal ordering. We must emphasise that any such
    theory based on many-body transformation would suffer from the unavoidable drawback.
    
    \item While it is absolutely necessary that one adopts normal ordering and eliminates the $S-S$ contractions
    in order to generate a terminating series of the effective Hamiltonian, this leaves
    out the clustering effects, which would otherwise be present in a conventional (full) exponential structure with $S$.
    
    \item The conventional iCCSDn, in practice, relies on the
    choice of a set of chemically active orbitals. The contractible (destruction)
    operators ($m$ and $e$) are restricted to only those
    active orbitals. The cost associated with the determination of the $s-$amplitudes is usually dictated by the construction of the $(h^{eff})_{ie}^{ab}$ residue, which formally scales as $n_on_v^4n_v^{act}$. Here, $n_o$ is the number of electrons, $n_v=(N-n_o)$ is the number of virtual spinorbitals and $N$ is the total number of spinorbitals. 
    $n_v^{act}=(N-n_o)^{act}$ 
    designate the number of chemically active virtual spinorbitals (the dimension of $e$).
    This, in practice, is less than the $n_o^2n_v^4$ which is the typical scaling 
    of CCSD, and thus the most expensive step is still dictated by the conventional CCSD
    terms. One may, however, choose to expand the choice of the active space to include 
    more occupied and/or virtual orbitals to include more complete description of 
    the connected triple excitations at the cost of an increased computational scaling. 
    Note that in iCCSDn, 
    the $T$ and the $S$ operators are simulated through a set of \textit{local denominators}, which can be perturbatively estimated by solving $[H_0, S^{(1)}] + V = 0$. Solution to this results in the following perturbative structure of the $s-$amplitudes:
\begin{equation}
\begin{split}
(-\epsilon_{i}+\epsilon_{a}+\epsilon_{b}-\epsilon_{e})
(s_p)^{ab}_{ie} + v^{ab}_{ie} & =0 \\
\Rightarrow (s_p)^{ab}_{ie} = 
v^{ab}_{ie}/(\epsilon_{i}+\epsilon_{e}-\epsilon_{a}- \epsilon_{b})
\label{eqxx19}
\end{split}
\end{equation}  
and 
\begin{equation}
(s_h)_{ij}^{am}=v_{ij}^{am}/(\epsilon_{i}+\epsilon_{j}-
\epsilon_{a}-\epsilon_{m}), 
\label{eqyy19}
\end{equation}
where $\epsilon_p$ is the orbital energy of the $p^{th}$ spinorbital.
    Thus the local energy denominator involves energies of only those orbitals 
    that appear in a given $S$ operator, and is equal to the energy difference between the
    doubly excited determinant and the triply excited determinant. While inclusion of a 
    large number of `active' orbitals generally improves the quality of the 
    results, it also makes the theory more prone to intruders, particularly in the 
    regions of molecular strong correlation. Thus, there is a delicate balance 
    between the accuracy and computational scaling that one needs to maintain 
    in order to solve the energetics of molecular systems in an uniform manner.
    
\end{enumerate}

In order to overcome the potential drawbacks in its implementation in a classical 
computer as described above, it is warranted that
the full exponential structure of $\exp(S)$ is utilized such that the $S$ operators 
are allowed to contract among themselves. However, in such a case, the many-body
first similarity transformed effective Hamiltonian $W$ would
generate exponentially large number of algebraic terms with all possible many-body ranks,
which are intractable to handle on a classical computer. Also the amplitudes during the 
optimization process are likely to be highly
plagued by the intruders, as described in point 3 above. Thus, rather
than taking the many-body route, one may evolve starting from 
the HF reference determinant in a quantum device by means of 
a sequential evolution through a fixed structure double unitary
ans\"{a}tz and variationally optimize the energy to get the updated set of parameters.

\subsubsection{Development of the unitarized iCCSDn ans\"{a}tz and implementation}

Towards the development of a unitary variant of iCCSDn (to be referred to as UiCCSDn 
henceforth) for its implementation in the VQE framework, the normal ordering of 
$e^S$ is removed by the conventional exponential structure and the $S$ operators are 
replaced by an anti-Hermitian sum of $S$ and $S^\dagger$. Similarly, the $T$ operators are 
replaced by the anti-Hermitian sum of the excitation
operator ($T$ itself) and the de-excitation operators ($T^\dagger$):
\begin{eqnarray}
& S_h \rightarrow S_h - S_h^\dagger = \sigma_h : {(\sigma_h)}_{ij}^{am} = {(S_h)}_{ij}^{am} - {(S_h)}_{am}^{ij} \nonumber \\
& S_p \rightarrow S_p - S_p^\dagger = \sigma_p : {(\sigma_p)}_{ie}^{ab} = {(S_p)}_{ie}^{ab} - {(S_p)}_{ab}^{ie} \nonumber \\
& T_2 \rightarrow T_2 - T_2^\dagger = \tau_2 : \tau_{ij}^{ab} = T_{ij}^{ab} - T_{ab}^{ij} \nonumber \\
& T_1 \rightarrow T_1 - T_1^\dagger = \tau_1 : \tau_{i}^{a} = T_{i}^{a} - T_{a}^{i}.  
\end{eqnarray}
Thus, in Eq. (\ref{ansatzeq}), both the exponential terms can be replaced by their 
unitary counterparts:
\begin{equation}
    U = U_1 U_2 = \exp{(\sigma)} \exp{(\tau)}, 
\end{equation}
where $\sigma = \sigma_h + \sigma_p$ and $\tau=\tau_1 + \tau_2$. Note that due to the 
exponential structure, the effective Hamiltonian can in principle be expanded in the 
Baker-Campbell-Hausdorff expansion to obtain normalized Hamiltonian expectation value
with variational optimization. However, due to the contraction among various components
of the ans\"{a}tz, the BCH expansion is non-terminating, making it intractable on a  classical computer.

The unitary version of iCCSDn may be employed to study the coherent time evolution 
in a quantum computer. Towards this, the ans\"{a}tz is broken down into several time
ordered sequences of few-particle operators using Trotter expansion. One may note 
that the $S$ and the $T$ operators do not commute in general. Since we have started 
with a double unitary structure and the $\sigma$ and $\tau$ operators do not commute, 
$e^\sigma e^\tau \ne e^{\sigma+\tau}$. However, from each 
set of operators, one may group terms which are mutually commuting such that 
\begin{eqnarray}
%&    e^{\sigma_A + \sigma_B} e^{\tau_A + \tau_B} = \nonumber \\ & \left(\lim_{\rho\to \infty} (e^{\sigma_A/\rho}e^{\sigma_B/\rho})^\rho\right) \left( \lim_{\lambda\to \infty} (e^{\tau_A/\lambda}e^{\tau_B/\lambda})^\lambda\right)
e^{\sigma_A + \sigma_B} e^{\tau_A + \tau_B} =   \left(\lim_{\rho\to \infty} (e^{\sigma_A/\rho}e^{\sigma_B/\rho})^\rho\right)\times \nonumber \\ \left( \lim_{\lambda\to \infty} (e^{\tau_A/\lambda}e^{\tau_B/\lambda})^\lambda\right).
\end{eqnarray}
Although a truncated Trotter expansion of the set of non-commuting operators 
is an approximation in the UiCCSDn wavefunction, the choice of the operators
provide us with enough variational flexibility to simulate strongly correlated 
systems. Even with $\rho=\lambda=1$ 
and a suitably approximated $\sigma$ operator, it can reproduce results which
are well within the limit of chemical accuracy, irrespective of the degree
of correlation of the system. The Trotterized ans\"{a}tz thus takes a form:
\begin{equation}
    | \Psi_{UiCCSDn}\rangle = \left( \prod_{q\in\{q_h, q_p\}} e^{\sigma_q} \prod_{r \in \{q_1, q_2\}} e^{\tau_r}\right)|\Psi_{HF}\rangle. 
\end{equation}
Here, $q_h$ and $q_p$ are the unique quartet of all $(i,j,a,m)$ and $(i,e,a,b)$ 
respectively, and $r$ takes the unique combinations of all singles and doubles.
This is realized through a concatenated quantum circuit where $e^\tau$ (with
selection of $\tau$ operators in lexical ordering) acts on the HF state first, 
followed by $e^{\sigma_h}$ and $e^{\sigma_p}$ 
(with selection of the $\sigma_h$ and $\sigma_p$ operators in lexical
ordering among themselves). One may further increase the variational flexibility 
of the ans\"{a}tz by independently taking a product of $k$ and $l$ number of $e^{\sigma}$ 
and $e^{\tau}$ unitary circuits, respectively. The ans\"{a}tz thus takes a generalized form:
\begin{eqnarray}
    %&   |\Psi_{UiCCSDn}\rangle =\nonumber \\
    %&\left(\prod_{\alpha=1}^k \prod_{q\in\{q_h, q_p\}} e^{\sigma_q^{(\alpha)}}\right)\left( \prod_{\beta=1}^l \prod_{r \in \{q_1, q_2\}} e^{\tau_r^{(\beta)}}\right)|\psi_{HF}\rangle. \nonumber 
    |\Psi_{UiCCSDn}\rangle=\left(\prod_{\alpha=1}^k
    \prod_{q\in\{q_h, q_p\}} e^{\sigma_q^{(\alpha)}}\right)\times \nonumber\\
    \left( \prod_{\beta=1}^l \prod_{r \in \{q_1, q_2\}} e^{\tau_r^{(\beta)}}\right)|\Psi_{HF}\rangle. 
\end{eqnarray}
Here $\sigma_q^{(\alpha)}$ and $\tau_r^{(\beta)}$ are treated as independent
parameters. We will show that for a few specific variants, the circuit for 
$e^{\sigma}$ is quite shallow \textit{(vide infra)}, allowing us to
increase the value of $k$ (keeping $l=1$) without significant increase in the
circuit depth. A schematic representation of the quantum circuit is shown in Fig. 
\ref{circuit} below.

One may note that choosing the ans\"{a}tz through a double unitary has an added 
advantage over a single unitary where one may write the unitary evolution
operator as $e^{\sigma + \tau}$. In the first order Trotter 
decomposition, although both ways of writing the ans\"{a}tz may superficially 
look similar, in our double unitary case, the order of the action of 
$e^\sigma$ and $e^\tau$ are fixed, contrary to a
single unitary. For the single unitary ans\"{a}tz like $e^{\sigma + \tau}$, 
one needs to optimize the order of appearance of various terms in the circuit. 
Typically, those operators should be allowed to act on the HF state first which do not
annihilate the reference. In our case, $e^\sigma$, by construction, is allowed
to act on the entangled state, $e^\tau | \Psi_{HF} \rangle$, It was found that the energy depends on the ordering of the operators even among the two separate circuits; however, in 
the current study both $\tau$ and $\sigma$ operators are 
taken in their lexical ordering. 
%to be 
%sufficiently resilient to the ordering among each group of operators, 
%even in the case of strongly correlated systems. 
One may also note that our ans\"{a}tz allows us to repeat a
part of the whole circuit to improve accuracy in a 
systematic manner.

\begin{table*}[t]
\renewcommand{\arraystretch}{1.0}
\begin{tabular}{|l|p{3cm}|p{8cm} |}
\hline
  Variants  & Choice of $\sigma$ & Description\\
\hline
\hline
\textbf{UiCCSDn} &  $\{{(\sigma_h)}_{ij}^{am}; (\sigma_p)_{ie}^{ab}\}$  & 
    1. Untruncated.

\\  
\hline
\textbf{UiCCSDn-A} &  $\{{(\sigma_h)}_{ij}^{au}; (\sigma_p)_{iv}^{ab}\}$  & 
    1. Active orbitals. 

\\  
\hline
  \textbf{UiCCSDn-A-op} &  $\{{(\sigma_h)}_{i_{\alpha}j_{\beta}}^{a_{\alpha}u_{\beta}},
{(\sigma_h)}_{i_{\beta}j_{\alpha}}^{a_{\beta}u_{\alpha}};$
\newline
${(\sigma_p)}_{i_{\alpha}v_{\beta}}^{a_{\alpha}b_{\beta}},
{(\sigma_p)}_{i_{\beta}v_{\alpha}}^{a_{\beta}b_{\alpha}}
\}$ & 

1. Active orbitals.
\newline
2. Opposite spin in scattering and excitation vertex of $\sigma$.

\\
\hline
\textbf{UiCCSDn-A-diag} & $ \{{(\sigma_h)}_{iu}^{au}; {(\sigma_p)}_{iv}^{av}\} $ & 

1. Active orbitals.
\newline
2. Diagonal terms in scattering vertex of $\sigma$.
\\
 \hline
  \textbf{UiCCSDn-A-pp} &
  $
\{{(\sigma_h)}_{i_{\alpha}i_{\beta}}^{a_{\alpha}u_{\beta}},
{(\sigma_h)}_{i_{\beta}i_{\alpha}}^{a_{\beta}u_{\alpha}};$
\newline
${(\sigma_p)}_{i_{\alpha}v_{\beta}}^{a_{\alpha}a_{\beta}},
{(\sigma_p)}_{i_{\beta}v_{\alpha}}^{a_{\beta}a_{\alpha}}
\}$ &

1. Active orbitals.
\newline
2. Same spatial orbitals for quasi-hole (in $\sigma_h$) and quasi-particle (in $\sigma_p$).
\\
%\hline
\hline
\end{tabular}
\caption{Summary of various approximate UiCCSDn schemes depending on the choice of the $\sigma$ operators. The cluster operator $\tau$ is taken to be all $1h-1p$ and $2h-2p$ excitations without any further truncation.}
\label{tab:variants}
\end{table*}

\subsubsection{Various approximate UiCCSDn schemes and the associated resource estimation}

\begin{table*}[t]
\renewcommand{\arraystretch}{1.5}
\begin{tabular}{|c|c|c|c|c|c|c|}

\hline
  Ans\"{a}tz  & $T_{1}$ & $T_{2}$ & $T_{3}$ & $S_{h}$ & $S_{p}$ & Gate depth \\
\hline
\hline
  UCCSD & $\mathcal{O}(n_{o}n_{v})$ & \cellcolor{lightgray}\boldmath{$\mathcal{O}(n_{o}^{2}n_{v}^{2})$} & - & - & - & $\mathcal{O}(n_{o}n_{v}^{2})$ \\

UCCSDT & $\mathcal{O}(n_{o}n_{v})$ & $\mathcal{O}(n_{o}^{2}n_{v}^{2})$ & \cellcolor{lightgray}\boldmath{$\mathcal{O}(n_{o}^{3}n_{v}^{3})$} & - & -  & $\mathcal{O}(n_{o}^{2}n_{v}^{3})$\\
UiCCSDn & $\mathcal{O}(n_{o}n_{v})$ & $\mathcal{O}(n_{o}^{2}n_{v}^{2})$ & - & $\mathcal{O}(n_{o}^{3}n_{v})$  & \cellcolor{lightgray}\boldmath{$\mathcal{O}(n_{o}n_{v}^{3})$} & $\mathcal{O}(n_{v}^{3})$    \\

UiCCSDn-A & $\mathcal{O}(n_{o}n_{v})$ & \cellcolor{lightgray}\boldmath{$\mathcal{O}(n_{o}^{2}n_{v}^{2})$} & - & $\mathcal{O}(n_{o}^{2}n_{v}n_{o}^{act})$  & $\mathcal{O}(n_{o}n_{v}^{2}n_{v}^{act})$ & $\mathcal{O}(n_{o}n_{v}^{2})$   \\

UiCCSDn-A-op & $\mathcal{O}(n_{o}n_{v})$ & \cellcolor{lightgray}\boldmath{$\mathcal{O}(n_{o}^{2}n_{v}^{2})$} & - & $\mathcal{O}(\frac{1}{16}n_{o}^{2}n_{v}n_{o}^{act})$  & $\mathcal{O}(\frac{1}{16}n_{o}n_{v}^{2}n_{v}^{act})$ & $\mathcal{O}(n_{o}n_{v}^{2})$   \\

UiCCSDn-A-diag & $\mathcal{O}(n_{o}n_{v})$ & \cellcolor{lightgray}\boldmath{$\mathcal{O}(n_{o}^{2}n_{v}^{2})$} & - & $\mathcal{O}(n_{o}n_{v}n_{o}^{act})$  & $\mathcal{O}(n_{o}n_{v}n_{v}^{act})$ & $\mathcal{O}(n_{o}n_{v}^{2})$   \\

UiCCSDn-A-pp & $\mathcal{O}(n_{o}n_{v})$ & \cellcolor{lightgray}\boldmath{$\mathcal{O}(n_{o}^{2}n_{v}^{2})$} & - & $\mathcal{O}(\frac{1}{8}n_{o}n_{v}n_{o}^{act})$  & $\mathcal{O}(\frac{1}{8}n_{o}n_{v}n_{v}^{act})$   & $\mathcal{O}(n_{o}n_{v}^{2})$\\

\hline
\end{tabular}
\caption{The number of parameters (and gates) required for each class of operators for the implementation of UCCSD, UCCSDT and various UiCCSDn variants. The leading gate count for each of the ans\"{a}tz are highlighted and put in the bold text. Also the corresponding leading gate depth is reported.
Note that UiCCSDn variants have much fewer number of gates required than UCCSDT.}
\label{tab:resources}
\end{table*}

In this section, we will introduce a number of variants of UiCCSDn ans\"{a}tz, in which we 
will judiciously approximate the $S$ operators such that the resulting ans\"{a}tze
strikes the right balance between expressibility, accuracy, number of 
parameters and circuit depth. We will also justify our choice from the 
perspective of many-body perturbation theory. Noting the fact that our scheme is 
not agnostic to the choice of holes and particles, we comment that the introduction 
of an operator with structures like that of $S$ is the minimal
requirement to introduce connected high-rank correlation effects beyond singles
and doubles, and thus the resulting ans\"{a}tz is expected to bring in better span of the Hilbert space. Furthermore,
one may significantly cut down the number of free parameters (and gate count)
by restricting the contractible orbitals (which appears as the quasi-hole/particle 
destruction operator) to some chemically \textit{active} ones. We introduce the
variants below which differ from each other in the 
choice of the scattering component in $\sigma$. Please note that in all the 
variants, the hole-particle structure of $\tau$ is retained in full.

\begin{enumerate}
    \item \textbf{UiCCSDn:} In the parent UiCCSDn theory, both the $\sigma$ and $\tau$ 
operators are retained in full and there is no further approximation involved. 
That implies that the contractible set of orbitals
$m$ and $e$ in $\sigma_h$ and $\sigma_p$ respectively are taken to be all 
the occupied and unoccupied orbitals. Implementation of the parent (untruncated)
UiCCSDn requires $\mathcal{O}(n_{o}^{3}n_{v})$ $\sigma_h$ and
$\mathcal{O}(n_{o}n_{v}^{3})$ $\sigma_p$ parameters on top of the conventional
$\mathcal{O}(n_{o}^2n_{v}^{2})$ elements of $\tau$. Note that the leading order of 
parameter count and gate count are dictated by the number of $\sigma_p$ elements,
while the circuit depth scales as $\mathcal{O}(n_{v}^{3})$.

\item \textbf{UiCCSDn-A:} In this variant, the contractible (hole and particle)
orbital indices are restricted to the chemically active orbitals. Since in our ans\"{a}tz, 
the high-rank correlation is simulated through the contraction of the $S$ and the $T$
operators, we span the full $N-$electron Hilbert space through only those $S$ operators
which involve contractible orbitals near the Fermi level. Note that in this variant,
there is no loss in the expressibility of the ans\"{a}tz as it spans the full $N-$electron
Hilbert space; however, they are generated through a selected contraction of the most 
dominant $\sigma$ with the $\tau$ amplitudes.
\begin{eqnarray}\label{Sactive}
\sigma_{\scaleto{UiCCSDn-A}{3.6pt}} \in \{{(\sigma_h)}_{ij}^{au}; (\sigma_p)_{iv}^{ab}\}
\end{eqnarray}
where $u$ and $v$ are the ``active" hole and particle orbital lines, respectively.
Noting the fact that $n_o^{act}, n_v^{act} << n_o, n_v$, the parameter count is
$\mathcal{O}(n_{o}^2n_{v}^{2})$ due to $\tau$. The circuit depth scales 
$\mathcal{O}(n_{o}n_{v}^{2})$.

\item \textbf{UiCCSDn-A-op:} On top of the selection of the active orbitals as
the contractible indices, we only incorporate those $\sigma$ operators in which 
the spins in the excitation vertex and the scattering vertex of $\sigma$ are 
different. This implies that some of the high-spin correlation channels are 
switched off.
\begin{eqnarray}\label{S_no_same_spin}
\sigma_{\scaleto{UiCCSDn-A-op}{3.6pt}} \in \{{(\sigma_h)}_{i_{\alpha}j_{\beta}}^{a_{\alpha}u_{\beta}},
{(\sigma_h)}_{i_{\beta}j_{\alpha}}^{a_{\beta}u_{\alpha}}; \nonumber \\
{(\sigma_p)}_{i_{\alpha}v_{\beta}}^{a_{\alpha}b_{\beta}},
{(\sigma_p)}_{i_{\beta}v_{\alpha}}^{a_{\beta}b_{\alpha}}
\}
\end{eqnarray}  
$u$ and $v$ denote active spatial occupied and virtual orbitals, and 
$\alpha$ and $\beta$ denote the spin-up and spin-down electrons, respectively.
While the number of $\sigma_h$ and $\sigma_p$ parameters in the UiCCSDn-A-op
variant scale similar to UiCCSDn-A, there is a small pre-factor of $1/16$ in the
former case. Thus for practical applications of UiCCSDn-A-op, there is a 
tremendous reduction in the number of the free parameters over the UiCCSDn-A
variant. While this requires $\mathcal{O}(\frac{1}{16}n_{o}^{2}n_{v}n_{o}^{act})$ 
number of $\sigma_h$ and $\mathcal{O}(\frac{1}{16}n_{o}n_{v}^{2}n_{v}^{act})$
number of $\sigma_p$ parameters, the parameter count is, however, dictated 
by $\mathcal{O}(n_{o}^2n_{v}^{2})$ number of $\tau$ amplitudes with a circuit 
depth of $\mathcal{O}(n_{o}n_{v}^{2})$.

\item \textbf{UiCCSDn-A-diag:} Note that the orbitals involved in the scattering
vertex of $S$ appear with opposite signs in their perturbative energy denominator
(see Eqs. (\ref{eqxx19}) and (\ref{eqyy19})). 
One may thus decipher that the most dominant terms in $S$ (or $\sigma$) are those
which have diagonal scattering. In the UiCCSDn-A-diag variant, we thus 
restrict the orbitals involved in the scattering vertex of $\sigma$ to be the same.
Inclusion of diagonal terms in the scattering vertex reduces the computation 
cost of Jordan-Wigner overhead significantly.
\begin{eqnarray}\label{Sdiag}
\sigma_{\scaleto{UiCCSDn-A-diag}{3.6pt}} \in \{{(\sigma_h)}_{iu}^{au}; {(\sigma_p)}_{iv}^{av}\}
\end{eqnarray}
The number of $\sigma_h$ and $\sigma_p$ parameters are further reduced to 
$\mathcal{O}(n_{o}n_{v}n_{o}^{act})$ number of $\sigma_h$ and
$\mathcal{O}(n_{o}n_{v}n_{v}^{act})$ number of $\sigma_p$ elements. The leading
parameter count thus still scales as $\mathcal{O}(n_{o}^2n_{v}^{2})$ with a
circuit depth of $\mathcal{O}(n_{o}n_{v}^{2})$.

\item \textbf{UiCCSDn-A-pp:} In the UiCCSDn-A-pp variant (where ``\textit{pp}" stands for 
partial pairing), in addition to the
active orbital selection in the destruction component of $S$, we include 
(i) only those quasi-hole creation operators which originate from the same 
spatial orbitals (for $\sigma_h$) and (ii) only those quasi-particle creation 
operators which have the same spatial orbitals (for $\sigma_h$). 
\begin{eqnarray}
\sigma_{\scaleto{UiCCSDn-A-pp}{3.6pt}} \in 
\{{(\sigma_h)}_{i_{\alpha}i_{\beta}}^{a_{\alpha}u_{\beta}},
{(\sigma_h)}_{i_{\beta}i_{\alpha}}^{a_{\beta}u_{\alpha}}; \nonumber \\
{(\sigma_p)}_{i_{\alpha}v_{\beta}}^{a_{\alpha}a_{\beta}},
{(\sigma_p)}_{i_{\beta}v_{\alpha}}^{a_{\beta}a_{\alpha}}
\}
\end{eqnarray}
Note that there is only partial pairing either among the quasi-hole or quasi-particle 
orbitals and the structure of the $\sigma$ operators does not allow to have complete 
pairing.
This variant further lowers the number of $\sigma_h$ and $\sigma_p$ elements
over UiCCSDn-A-diag and the gate count and the circuit depth scales similar to 
UiCCSDn-A-diag. 
\end{enumerate}

One may note that in all the variants, except the parent (untruncated) 
UiCCSDn, there is no increase in the order of the gate count and circuit depth
over the conventional UCCSD. However, we will demonstrate that all our variants 
predict energy well within the chemical accuracy.
A succinct summary of the choice of the various operators and the associated parameter 
count in different approximate schemes are given in Table. \ref{tab:variants} and \ref{tab:resources} 
respectively.

\section{Results and discussions}\label{results}

\subsection{Methodology}

\begin{figure}[t]
\includegraphics[height=65mm,width=85mm]{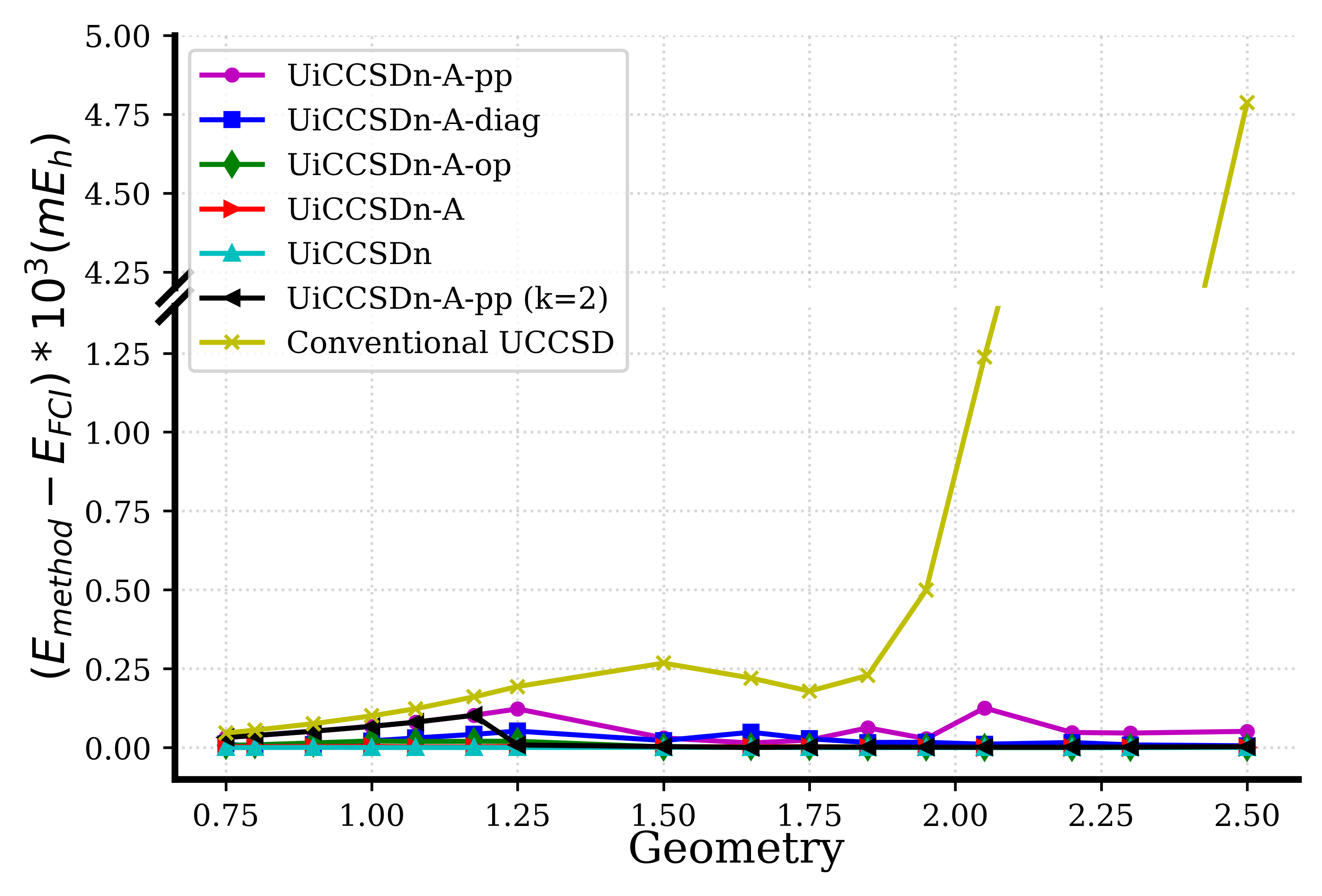}
\caption{Difference between VQE energies corresponding to various ans\"{a}tze and FCI energies for $H_{2}O$ at symmetric stretching in STO-3G basis.}
\label{one}
\end{figure}

\begin{figure}[t]
\centering
\includegraphics[height=75mm,width=85mm]{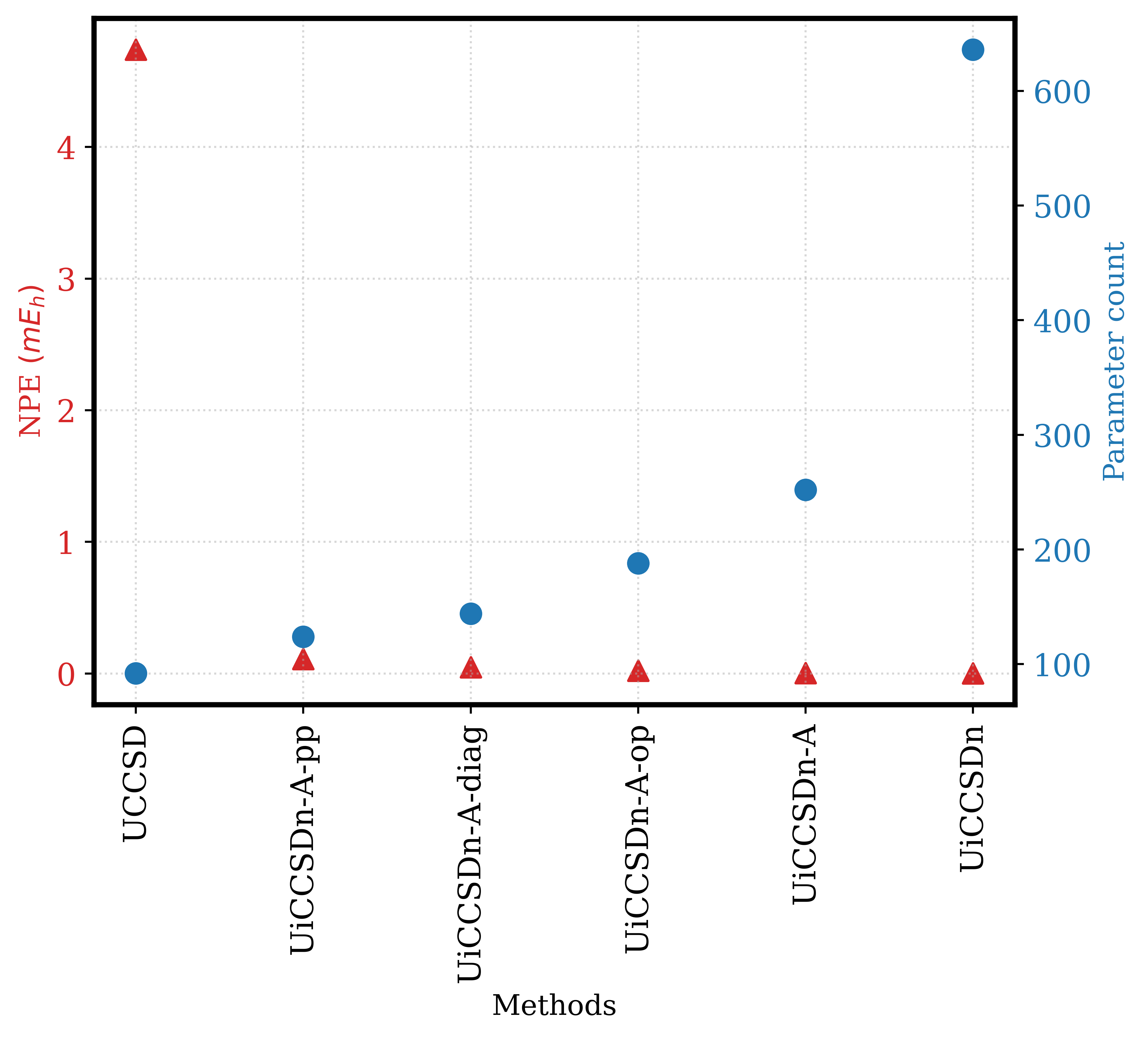} 
\caption{Non-parallelity error (red triangles) and parameters count (blue dots) corresponding to various ans\"{a}tze for the symmetric stretching of $H_{2}O$ in STO-3G basis.}
\label{two}
\end{figure}

\begin{table*}[t]
\centering\setlength{\extrarowheight}{2pt}
%\centering
%\rowcolors{1}{gray}
%\rowcolors{2}{gray!10}{gray!40}
\begin{tabular}{|*{5}{p{2.7cm}|}}
\hline
 \multirowcell{2}{Method} & \multicolumn{4}{c|}{Energy ($E_{h}$)} \\
 \cline{2-5}
& \makecell{d=1\AA} & \makecell{d=2\AA} & \makecell{d=3\AA}& \makecell{d=4\AA}  \\
\hline
\hline
FCI & \makecell{-7.78446028} & \makecell{-7.86108777} & \makecell{-7.79884316} &\makecell{-7.78427817} \\
 & & & &\\[-2.5ex]
UCCSD  & \makecell{{-7.78445508}}\newline \makecell{\footnotesize{(0.0052)}} & \makecell{-7.86106905}\newline\makecell{\footnotesize{(0.0187)}} & \makecell{{-7.79875361}}\newline\makecell{{\footnotesize{(0.0895)}}} &\makecell{{-7.78414449}}\newline\makecell{{\footnotesize{(0.1336)}}} \\
& & & &\\[-2.5ex]
UiCCSDn-A-pp & \makecell{-7.78445982}\newline\makecell{{\footnotesize{(0.0004)}}} & \makecell{-7.86108722}\newline\makecell{{\footnotesize{(0.0005)}}} & \makecell{-7.79884258}\newline\makecell{{\footnotesize{(0.0005)}}} &\makecell{-7.78427777}\newline\makecell{{\footnotesize{(0.0004)}}} \\
& & & &\\[-2.5ex]
UiCCSDn-A-diag & \makecell{-7.78446017}\newline\makecell{{\footnotesize{(0.0001)}}} & \makecell{-7.86108736}\newline\makecell{{\footnotesize{(0.0004)}}} & \makecell{-7.79884162}\newline\makecell{{\footnotesize{(0.0015)}}} &\makecell{-7.78427446}\newline\makecell{{\footnotesize{(0.0037)}}} \\
& & & &\\[-2.5ex]
UiCCSDn-A-op & \makecell{-7.78446000}\newline\makecell{{\footnotesize{(0.0002)}}} & \makecell{-7.86108766}\newline\makecell{{\footnotesize{(0.0001)}}} & \makecell{-7.79884256}\newline\makecell{{\footnotesize{(0.0006)}}} &\makecell{-7.78427780}\newline\makecell{{\footnotesize{(0.0003)}}} \\
& & & &\\[-2.5ex]
UiCCSDn-A & \makecell{-7.78446025}\newline\makecell{{\footnotesize{(0.0000)}}} & \makecell{-7.86108772}\newline\makecell{{\footnotesize{(0.0000)}}} & \makecell{-7.79884302}\newline\makecell{{\footnotesize{(0.0001)}}} &\makecell{-7.78427781}\newline\makecell{{\footnotesize{(0.0003)}}} \\
\hline 
\end{tabular}
\caption{The energy values (in $E_h$) for some selected geometries of $LiH$ molecule using UCCSD and different UiCCSDn variants. The corresponding FCI energy values are also reported to demonstrate the accuracy of our schemes. The quantities in parenthesis denote the energy difference (in $mE_h$) from the FCI values. Note that all the energy values reported here used $k=l=1$.}
\label{tab:LiH}
\end{table*}

We have chosen to study the potential energy surface for symmetric stretching of $H_{2}O$ 
and LiH to assess the performance of the newly developed UiCCSDn ans\"{a}tz and the subsequent 
variants in the context of VQE. For all our computations, we have chosen the contracted STO-3G 
basis~\cite{sto}.
The one- and two- electron integrals were taken from PySCF~\cite{pyscf}, 
and the UiCCSDn ans\"{a}tz has been implemented in Qiskit
0.26~\cite{qiskit}. Also, the UCCSD calculations were done
using the same version of Qiskit whereas the UCCSDT calculations were
carried out in Qiskit nature. The Jordan-Wigner scheme~\cite{map}
was chosen for encoding the creation and annihilation operators into the
qubit operators. Also, we had employed the direct spin-orbital to qubit
mapping: that is each spin-orbital is represented as one qubit, 
independent of the occupancy of the
spin-orbitals. For the classical subpart of the VQE, we chose L\_BFGS\_B optimizer~\cite{L_BFGS_B} throughout. For all the calculations, the initial values of all the 
parameters, $\theta_{init}$, were set to zero. Further, in the present study, we have worked with the statevector backend simulator. The 
Trotter number for all our calculations was set to be one. 
%\begin{figure*}[!t]
%\includegraphics[height=65mm,width=85mm]{h2o.png} 
%\caption{Difference between VQE energies corresponding to various ans\"{a}tze and FCI %energies for $H_{2}O$ at symmetric stretching in STO-3G basis.}
%\label{one}
%\end{figure*}

\subsection{Symmetric stretching of $H_{2}O$}
\subsubsection{The potential energy surface and the non-parallelity error}

We determined the VQE energies employing our ans\"{a}tze for all the geometries along the potential energy surface. We have varied the O-H bond length from around 0.75 times the equilibrium distance to 2.5 times the equilibrium distance in order to generate the potential energy surface. For the purpose of comparison, we have also computed the VQE energies using the conventional UCCSD ans\"{a}tz. As shown in Fig. ~\ref{one}, we have plotted the differences between the VQE energies computed using the aforementioned ans\"{a}tze and the FCI energies. It is quite evident from the plot that VQE with the conventional UCCSD ans\"{a}tz performs quite well and agrees to 0.18 milliHartree ($mE_{h}$) with respect to FCI, for geometries below 1.75 times the equilibrium length, but starts deviating beyond that point. This trend shows that for the domains where strong correlation effects come into play, UCCSD fails to estimate the ground state energies. Whereas the energies obtained using the UiCCSDn-A-pp ans\"{a}tz (which is the least parametrized ans\"{a}tz amongst our other ans\"{a}tze) are precise up to tens of microHartree ($\sim 10 \mu E_h$) throughout the potential energy surface. In order to explicate our observations, we have further calculated the corresponding non-parallelity error, abbreviated as NPE. NPE is defined as the difference between the maximum and minimum deviations from the exact ground state energy, i.e., FCI in our case. We observe that the NPE corresponding to the conventional UCCSD and UCCSDT ans\"{a}tz is 4.74 $mE_{h}$ and 2.169 $mE_{h}$ respectively. On the other hand, the  NPE associated with the least parametrized ans\"{a}tz, UiCCSDn-A-pp,  is found to be 0.11  $mE_{h}$. The NPE and the parameters count associated with all the ans\"{a}tze have been plotted in Fig. \ref{two}. \\

In order to assess the efficiency of the newly developed ans\"{a}tze, it is important to
elucidate on the circuit complexity, which is quantified by the number of variational 
parameters and the quantum gate count. We should note that the NPE corresponding to 
UiCCSDn-A-pp is in the order of 0.11 $mE_{h}$ at the expense of an additional 32 parameters 
over the conventional UCCSD (with a NPE of 4.74 $mE_{h}$). Also it is worth mentioning at 
this point that the conventional UCCSDT leads to a NPE of 2.169 $mE_{h}$ at the expense of 
188 parameters (which is of the order $\sim\mathcal{O}(n_o^3n_v^3)$), whereas the UiCCSDn-A-diag (with lesser number of parameters than UCCSDT) 
and the UiCCSDn-A-op (with same number of parameters as UCCSDT) exhibit 0.048 $mE_{h}$ 
and 0.0209 $mE_{h}$ NPE respectively. Looking at the number of parameters (and hence, the  gate count), we can infer that the newly developed ans\"{a}tze strikes a right balance between the two aspects i.e., implementation cost and the accuracy. 
        
\subsection{Potential energy curve for $LiH$ dissociation}

We have evaluated the VQE energies employing different variants of UiCCSDn ans\"{a}tze 
for a few selected geometries of LiH, as listed in Table~\ref{tab:LiH}. According to our
findings, the conventional UCCSD approach performs pretty well around the equilibrium geometry, but it 
starts to deteriorate for the strongly correlated regions when the $Li-H$ bond is stretched.
However, the ground state energies evaluated through our ans\"{a}tze uniformly show a 
sub micro-hartree accuracy for both weakly and strongly correlated regions irrespective of 
the UiCCSDn variant. 

 \section{Conclusion and future outlook}\label{conclusion}
In this paper, we have developed the unitary version of the dual 
exponential coupled cluster methodology (UiCCSDn) and have adapted it for use in a quantum computing framework.
We have presented theoretical arguments for the non-viability of implementing the iCCSDn methodology or for that matter, its unitary version on a classical device without approximation in the rank and number of
terms in the similarity transformed many-body Hamiltonian. The implementation of UiCCSDn 
in a quantum device can bypass these drawbacks entirely. The dual exponential structure,
by construction, allows the scattering operators to act on entangled states to
induce high-rank excitations. We have introduced different variants
by judiciously including a selection of the scattering operators while keeping the cluster 
operators in full. Apart from the parent UiCCSDn, all the variants have a gate count 
and circuit depth in the same order as the conventional UCCSD without significantly 
compromising the expressibility of the anstze. We have shown with a prototypical 
molecular application that all the different variants show high accuracy compared to 
classically computed FCI over the entire molecular surface, but with a significantly
less implementation cost than allied theories.

It would be interesting to implement the UiCCSDn in the ADAPT-VQE framework to further 
reduce the parameter count. Evaluation of the parameters in PQE framework would be another
avenue which would keep us engaged in coming years.

\section{Acknowledgements}
RM thanks Science and Engineering Research Board (SERB),
Government of India for providing financial support. DH thanks
SERB and IRCC, IIT Bombay for research fellowship.
Some of the preliminary calculations were carried out on National Supercomputing Mission's (NSM) computing resource, `PARAM Siddhi-AI', at C-DAC Pune, which is implemented by C-DAC and supported by the Ministry of Electronics and Information Technology (MeitY) and Department of Science and Technology (DST), Government of India.

\section*{Data Availability}
The data generated in this study is available upon reasonable request to the corresponding author.

\end{document}